SCHOOL OF INFORMATION
UNIVERSITY OF MICHIGAN

# Globalization and Governance in Cyberspace:
## Mapping the Processes of Emergent Regime Formation in Global Information and Communications Policy

**Derrick L. Cogburn**
The University of Michigan
School of Information

**Comments are appreciated please send inquires to:**
Prof. Derrick L. Cogburn
301 West Hall, 550 East University Avenue
School of Information
The University of Michigan
Ann Arbor, Michigan, 48109
dcogburn@umich.edu

# Globalization and Governance in Cyberspace:
## Mapping the Processes of Emergent Regime Formation in Global Information and Communications Policy[1]

**Derrick L. Cogburn, Ph.D.**
The University of Michigan

## Abstract

This paper develops a theoretical perspective on globalization and the information society and combines it with a critical usage of international regime theory as a heuristic for understanding the current historical period of transition from an international telecommunications regime (Cowhey 1990, 1994) to a new and complex regime aimed at providing governance for the Global Information Infrastructure and Global Information Society (GII/GIS). In analyzing the principles, values, norms, rules, collective decision-making procedures, and enforcement mechanisms of the emergent GII/GIS regime, this paper differentiates between three regime levels: (1) Macro-Regime—global; (2) Mezzo-Regime—regional and sub-regional; and (3) Micro-Regime—national. The paper employs a case-study approach to explore some of the specific national responses (i.e. South Africa) to this regime transition, with an analysis of potential best practices and lessons learned for other emerging economies. Key findings in this paper are: (1) that a range of social, political, economic, and technological factors are eroding the existing international telecommunications regime (e.g., VOIP, call-back, VSATs, accounting rate restructuring, pressure for applications development, and SMMEs); (2) a new regime for global information and communications policy is emerging, but is being driven not by the broad possibilities of the Information Society, but by the more specific interests of global and multi-national corporations related to e-commerce; (3) numerous strategic responses have been developed at national, sub-regional, and regional levels to the challenges of this transition in both developed and developing regions; and (4) without a collaborative response, the developing world will be further marginalized by this new regime.

## Table of Contents

# **Globalization and Governance in Cyberspace:**
## Mapping the Processes of Emergent Regime Formation in Global Information and Communications Policy[2]

**Derrick L. Cogburn, Ph.D.**
The University of Michigan
School of Information

## **Introduction**

For more than three decades, international relations scholars have grappled with the question of the "anarchy problematique" and the evolution of cooperation at national, regional and global levels (see, *inter alia*, Keohane and Nye, 1977; Krasner, 1983; Keohane, 1984; Axelrod, 1984). These theorists have attempted to understand how global governance and cooperation can occur in a world-system of "sovereign and equal" national states, and in the absence of global government to make and enforce rules. International Regime theory has been one of the most resilient mental models for addressing this problem, and has been formulated from a wide variety of epistemological and scholarly traditions.

In 1983, a special issue of the journal *International Organization* attempted to build a consensus around the theoretical and applied approaches to International Regime theory. Here, Krasner (1983) and his colleagues defined regimes as "sets of implicit or explicit principles, norms, rules and decision-making procedures around which actors' expectations converge in a given areas of international relations." Principles are seen as beliefs of fact, causation or rectitude; Norms are seen as standards of behavior defined in terms of rights and oblications; Rules are specific prescriptions or proscriptions for action; and Decision-Making Procedures are the prevailing practices for making and implementing collective choice (Krasner 1983). More directly, international regimes are the "rules of the game" for a specific issue area within the world-system and the collective enforcement mechanisms for those rules.

Further, Rittberger (1995) argues that international regime theory, while once thought to be a "passing fad," has maintained exceptional stability and integrative capabilities within the discipline of international relations, and included the insights from international political economy, security experts, comparative politics, and other areas in world affairs. He asserts that regime theory has become an increasingly important intellectual tool in European scholarly circles, especially in Germany and Scandanavia, and attempts in his work to "enhance integration and communication" amongst scholars of international regimes (Rittberger, 1995).

Over these several decades, scholars around the world have documented the emergence and efficacy of International Regimes in a wide variety of issue areas within the world-system, including: (1) international shipping; (2) international air transport; (3) international post; (4) international atomic energy and weapons; (5) international environmental issues; (6) the global "commons" (e.g., the high seas and outer space); and (7) even for commodities (e.g. diamonds). However, one of the oldest and most successful international regimes has been the International Telecommunications Regime (Cowhey, 1990).

Based on the International Telecommunication Union (ITU) and an epistemological community that supported the concepts of the preferred natural monopoly for telecommunications, and the "clubby" and "cartelized" relationships between ministries and officials of monopoly Post, Telegraph and Telephone entities (PTTs), ITU officials (many of whom are former PTT officials), and a limited number of upstream suppliers to the PTT in the national capitals, the International Telecommunications Regime (ITR) was highly successful (Cowhey, 1990; Freiden, 1996). So successful was this regime, that it emerged before international regime theory was in vogue and has been largely unchallenged until recent times.

However, it is this regime that is now facing tremendous transformative pressures as the historical processes of globalization and an information age continue to unfold. A wide range of social, political, economic, cultural, and technological factors are challenging the International Telecommunications Regime and pressing for the emergence of a new regime, that might be called the Global Information Infrastructure/Global Information Society Regime (GII/GIS).

What are the emerging "principles, values, and norms," of this new regime, and what stakeholder interests are best represented by them? What will be the rules of this new regime, and in what international body will they be based? Who wins and who loses from this emerging regime? And, what can be done to influence the direction of this regime to ensure the development of a more just, and equitable Global Information Society? These are the research questions that this paper seeks to address.

## Nature of the Problem

### *Globalization, the Information Economy and Global Governance*

While there has been a tremendous amount written about Globalization and the information economy, there is still little consensus on the fundamental meaning of these two concepts, which are among the most contested of recent times. "Globalisation" means very different things to different people. The popular perception of globalization is simply that people, societies and nations are bound increasingly together, and there is a particular focus on how closely bound together are international markets. While this perception of interconnectedness is an important aspect of globalization, the actual processes go much further. McGrew (1992) asserts that globalization is:

> the multiplicity of linkages and interconnections that transcend the nation-states (and by implication societies) which make up the modern world system. It defines a process through which events, decisions, and activities in one part of the world can come to have significant

> consequences for individuals and communities in quite distant parts of the globe.[3]

One of the foremost thinkers about globalization and its processes has been Anthony Giddens. He considers globalization to be one of the most "visible consequences of modernity," and asserts that "globalization concerns the intersection of presence and absence, the interlacing of social events and social relations 'at a distance' with local contextualities" (Giddens).[4] Giddens also argues that "larger and larger numbers of people live in circumstances in which disembedded institutions, linking local practices with globalized social relations, organize major aspects of [their] day-to-day life" (Giddens).[5]

With this expanded and more precise definition, we see that globalisation is not just about the deepening of financial markets, but includes a whole range of social, political, economic, and cultural phenomena. We refer to these areas as the "spheres" of globalisation. One of the reasons that the Information Economy offers such promise to Africa is that each of these "spheres" of globalisation is supported by the application of electronic commerce. Also, through strategic planning, the opportunity exists for key geographic areas in Africa to exploit information and communications technologies to become "spaces" of globalisation.

*Fundamental Transformation in the Global Economy*

The information economy is based on a fundamental transformation of the underlying structure of the global political economy. Many analysts are now arguing that this change is so definitive that it warrants the label of a techno-economic paradigm shift. This shift reflects changes in science, technology, the organisation of business, production, learning and entertainment. Aspects of this transformation include: (1) the nature of the manufacturing company; (2) the changing nature of business dynamics; (3) major changes in the R&D activities of firms; (4) demand articulation in technological development; (5) technology fusion; and (6) institutional inertia. These changes are affecting nearly all sectors of the world-system, including intergovernmental organisations, the private sector, regional organisations, national states, and non-governmental organisations.

This period of change engenders a potential restructuring of power relations and the development of new forms of inequality in the world. It is possible that inequality in the information economy could go beyond a division between the so-called developed and developing countries to exacerbate intra-country divisions. Specifically, divisions could sharpen between those individuals possessing the knowledge, skills and abilities to contribute to the global information economy (wherever they may happen to be located) and those who do not possess such skills.

---

[3] Anthony McGrew, et al, eds. Global Politics Globalization and the Nation State,(Cambridge: Polity Press, 1992), 66.

[4] Anthony Giddens, Modernity and Self-Identity, 21.

[5] Ibid, 79.

*Towards a Definition of the Information Economy*

In defining the information economy, the author acknowledges the existence of a rich academic and popular debate on the subject. A key component of this debate, is the impact of Globalization on the formation of a "knowledge economy" in general, and the emergence of an "information economy" in particular. One characteristic of this current period of Globalization is the emergence of a new techno-economic paradigm, which some analysts call innovation-mediated production. Within this framework, knowledge is increasingly embedded within the production process itself. One major issue that contrasts the knowledge economy from the industrial economy is that in many cases, the barriers to entry are much lower. In the new economy, information and knowledge become the most important factors of production. This mode of production characterises the overall "knowledge economy," within which the "information economy" is playing an increasingly important role.

The author argues that the terms "information economy" and "knowledge economy" are very closely related and can be used synonymously in most cases. However, for the purposes of the paper, the author argues that the term "information economy" refers to a specific component of the emerging knowledge economy wherein the production of information goods and services dominates wealth and job creation.[6]

Perhaps the most important development within the information economy is the economic explosion caused by global electronic commerce (e-commerce). E-commerce is the production, distribution, marketing, sale or delivery of goods and services by electronic means. This includes the integrated use of information and communications technologies (e.g., the global information infrastructure) as the medium through which goods and services of economic value are researched, designed, produced, advertised, catalogued, inventoried, purchased, distributed, accounts settled, follow-up support provided, and management information systems implemented.

Already, e-commerce is facilitating a process of dis-intermediation, where traditional intermediary functions are being replaced by new products and intelligent agents. However, many new markets are being developed for information goods and to cope with such a dramatic increase in the availability of information, new information mediaries (infomediaries) are emerging. Whole new industries are emerging and new markets (and new types of markets) are being developed. Also, an unprecedented amount of information is being collected on individual consumers, allowing new and closer relationships to be forged between business and their customers, while at the same time creating new concerns about privacy in the on-line world.

The global information economy could be characterised as disciplinarian. Its interdependent nature ensures that “bad” decisions are punished immediately; and “good” decisions are rewarded with the same speed. With such a global, interdependent, knowledge-based economy, it is critical that appropriate mechanisms be developed at a global level to "govern" the global information economy—a global information economy regime.

[6] See for example, Shapiro & Varian (1998); Varian (2000), Tapscott (1997, 1998, 1999); Choi, et al (1997); Quah (2000); Kahin & Brynjolfsson (2000).

*Global and Regional Responses to the Information Revolution*

1st ITU World Telecommunications Development Conference (1994):

In response to these numerous challenges, countries and international organizations around the world have moved towards collective and individual efforts to harness resources, ideas and strategies. One of the earliest activities was the 1st World Telecommunications Development Conference (WTDC), hosted by the Development Bureau (BDT) of the International Telecommunications Union (ITU). Held in Buenos Aires, Argentina, the WTDC addressed the issue that was raised decades earlier in the Maitland Commission Report, chaired by Sir Donald Maitland. Known popularly as the "Missing Link" report, the Maitland Commission Report argued that there was a conclusive link between telecommunications penetration and socio-economic development. The WTDC reassessed this argument and ended with the same conclusions, adding that the gap between developing and developed countries had grown, not diminished, since the Missing Link. The conference attempted to harness the resources necessary to address this widening Gap.

Group of Seven (G7) and the Information Society (February, 1995):

Following this landmark meeting, ministers from the leading industrialized countries in the world gathered in Brussels, Belgium for the 1st Information Society Ministerial Meeting. These ministers were attempting to understand how they could work together to harness the increasing potential of information and communications technologies to address the increasing challenges facing their individual countries. Emerging from the G7 (actually the G8 with the inclusion of Russia at the meeting) Information Society ministerial meeting, was a series of Eleven Information Society pilot projects, that were designed to identify "best practices" and "lessons learned" from collaborative efforts amongst the G8 members. Several of these projects (namely the Government On-Line and the Global Marketplace for SMEs) actively encouraged participation from developing country members. Results from these pilot projects were analyzed continuously, and the European Commission established the Information Society Projects Office (ISPO) to further disseminate the lessons coming from the pilot projects

However, one major challenge for the meeting was that, as a G8 meeting, there was almost no participation from developing countries. The primary exception was that South African Deputy President Thabo Mbeki, was invited to speak "on behalf of the developing world." In his address, he argued that the participants at the meeting could not build a Global Information Society with only the eight participants sitting around the table, but that they had to involve a wide cross-section of the developing world. He offered South Africa as the host of such an initiative, and this challenge ultimately led to the G7/Developing World Information Society and Development Conference (ISAD) that will be discussed below.

Global Information Infrastructure Commission (February 1995):

While the G8 governments were meeting to develop their collaborative strategies for confronting the Information Society and exploring ways to involve the developing world, another important meeting was occurring in Brussels. A group of Chief Executive Officers from some of the leading information and communications companies in the world met to further challenge the G8 governments. As they launched their new organization, the Global Information Infrastructure Commission (GIIC), they argued that while Deputy President Mbeki was right in arguing for more developing country involvement in building the Information Society, it had to go even further. The GIIC members argued that governments alone would be unable to build the Information Society, and that private sector leadership was critical in partnership with the public sector if a truly "global" Information Society would be built. Their fifty members pledged to work for the next three years to help to promote the important role of the private sector in building the GII and to engage with public sector actors to help spread that message.

United Nations Economic Commission for Africa (1995/1996)

During the G7 ministerial meeting, Thabo Mbeki's remarks were well received. However, it took considerably longer for the G8 members to actually agree on a way to engage with the developing countries (in fact the agreement to participate in the ISAD conference only happened at their meeting in Novia Scotia). In the meantime, Africa did not wait for this potential blessing. In April 1995, the United Nations Economic Commission for Africa (ECA), hosted what has become known as a landmark meeting called the Telematics for African Development Symposium. The ECA is the largest UN presence in Africa. Headquartered in Addis Ababa, Ethiopia, it is one of three major regional organizations designed to facilitate regional socio-economic development in all 53 African countries (the other two being the Organization for African Unity, also located in Addis Ababa, and the African Development Bank, located in Cote D'Ivoire). This Telematics for African Development Symposium brought together numerous African experts in the use of information and communications technologies for development.

One result of the conference was a resolution for the Council of Ministers entitled "Building Africa's Information Highway" that called for an African response to the challenges of the Information Society. This resolution was adopted, and the Council created a High-Level Working Group (HLWG) on Information and Communications Technologies. Working mostly virtually (meeting physically only twice, once in Cairo and again Addis), the HLWG developed a high-level response for Africa called the African Information Society Initiative (AISI). The AISI was adopted by the Council of Ministers in May 1996 and was endorsed by the African Ministers of Communications meeting in Abidjan, Cote d'Ivoire to finalize the drafting of the *African Green Paper on Telecommunications.* One key component of the AISI is its focus on National Information and Communications Infrastructure (NICI) planning in each African country. AISI was given its public launch at in South Africa at a luncheon hosted by the GIIC at the Information Society and Development (ISAD) conference.

Information Society and Development Conference (May 1996):

As the result of Thabo Mbeki's challenge to the G8, the Information Society and Development Conference was hosted in South Africa in May 1996. As planned, a wide cross-section of developing countries participated in the ISAD conference, and challenged significantly the process of regime formation being led by the highly industrialized countries. Key issues placed on the global agenda were the following: (1) multi-purpose community information centers; (2) universal access; and (3) employment issues. Major plans to host a follow-up conference (ISAD II, proposed by the Arab Republic of Egypt) never materialized.

ITU Universal Right to communicate:

Moving in the same direction, Dr. Pekka Tarjanne, Secretary General of the International Telecommunication Union (ITU), proposed that another principle be added to the United Nations Universal Declaration of Human Rights; this being the Universal Right to Communicate. Several other organizations have now taken up this cry and are moving towards trying to include this principle on the larger global agenda. This addition elevates universal access from a luxury to a recognized basic human right.

Partnership for ICTs in Africa:

The AISI is grounded in the assumption that effective socio-economic development requires partnerships between many partners, including: (1) international development agencies; (2) donor agencies; (3) private sector actors; (4) non-governmental actors.

Global Knowledge for Development (June 1997):

In the vacuum left by the non-starting ISAD II, the World Bank initiated a similar conference to fill the void, but with a fairly different character. The Global Knowledge for Development conference was held in June 1997 in Toronto, Canada and organized by the World Bank and the government of Canada. The developed country orientation of this new conference was quite apparent. South African Minister of Communications Jay Naidoo, took the podium at one point and argued that "this conference is not the ISAD follow-up, that conference is yet to take place." Unfortunately, that conference has still not taken place (a second Global Knowledge conference was held in February 2000 in Malaysia, but there were significant confrontations between the developing and developed country contingents and international organizations).

2nd ITU World Telecommunications Development Conference (1998):

The tiny island of Malta played hosted to the second ITU World Telecommunications Development Conference in 1998. Attempting to assess the progress on the Buenos Aires Action Plan and to produce the program of work for the BDT for the next four years. Key issues gaining prominence at this conference was the addition to the agenda of a focus on women's access to information and communications technologies. The conference ended with the adoption of the Valetta Action Plan, hoping to consolidate the progress towards the development of an Information Society.

OECD Global Electronic Commerce Ministerial Meeting (1998):

In 1998, another distinctive shift began to happen in the movement towards an Information Society. While commercial issues have always been important, the increasing importance of Electronic Commerce (e-commerce) and a global information economy began to dominate discussions of an Information Society. In 1998, the Organization for Economic Cooperation and Development (OECD), which represents the 27 most industrialized countries of the world, hosted a ministerial level conference on global electronic commerce. A few developing countries were invited to participate in the event (including Minister Naidoo of South Africa), but many complained that their interests were clearly not important to the overall OECD agenda (the OECD has now planned a follow-up to this conference for July in the Middle East).

European Commission Information Society Technologies Conference (1998):

As an attempt to counter the perceived continued US dominance in information and communications technologies, the European Commission launched a new series of conferences for its members and strategic partners. These Information Society Technologies (IST) conferences started in 1998 in Vienna and were designed for focus on stimulating cooperative research through the 5th Framework Initiative for collaborative research.

Africa TELECOM (May 1998):

The International Telecommunication Union is primarily a politically-oriented inter-governmental organization. Its meetings are mostly intergovernmental negotiations on standards and agreements. However, it also has an alternative format for meeting called its TELECOM conferences. These conferences are able to raise issues and try to promote consensus in a less politically charged and threatening environment. World TELECOM is the centerpiece event, and it occurs every four years in Geneva, Switzerland. Given its quarter-annual format, it is often referred to as the “Olympics of telecommunications.” After the World event, the conferences move around the world to the three major regions (Asia TELECOM, Africa TELECOM and Americas TELECOM). In May 1998, South Africa hosted the Africa TELECOM event and brought considerable attention to the perspectives of Africa in the movement towards and Information Society. Also during the Africa TELECOM, the Global Information Infrastructure Commission launched its first regional organization called GIIC Africa. GIIC Africa, with its motto of “Africanizing the Global Information Society with Private and Public Sector Cooperation,” is designed to bring together 50 African private sector leaders to create additional momentum towards building the GII/GIS in Africa. Further, a grouping of African ministers of communications used the conference to issue its *African Connection* agenda, designed to further develop consensus amongst the African public sector on the importance of information and communications technologies to development.

*Governance of the Global Information Economy and Society*

A transformation of such historic proportions is engendering substantial change in the mechanisms of governance as well. The spatially disarticulated nature of the global economy and its research & development, production, testing, distribution, and management systems is challenging models of global governance. These are the issues that are leading the revolution in the International Telecommunications Regime, and that will influence the emergence of the new Global Information Infrastructure/ Global Information Society Regime (GII/GIS). In this case, the international regime of norms, principles, values and enforcement mechanisms for this new economy are being developed as various societal actors around the world attempt to influence this process. Further, as this new regime is being developed, many societal actors are assessing, and reassessing their roles and strategies.

Most likely, this new GII/GIS regime will be based upon the World Trade Organisation (WTO), the Geneva-based successor organisation to the General Agreement on Tariffs and Trade (GATT). Global market-access and a liberalised, rule-based, multilateral trading environment for tangible goods and intangible services are some of the key principles that already define this new economic order. These challenges require strategic responses from both state and non-state actors to include participation and partnerships from all relevant societal actors (public, private and voluntary sector actors). The senior public, private and civil-society leadership in Africa and other developing regions must find ways to strengthen their voices in the high-level processes of regime formation if the rules of the new economy are to adequately reflect some of their interests.

## Literature Review

There are three dominant schools in international regime theory: (1) liberal/neo-liberal; (2) realist/neo-realists and Marxist/neo-Marxist; (3) and what might be called (3) postmodernist. In the liberal/neo-liberal school, there is a focus on the importance of functions. Theorists working in this school, focus on the impact that international regimes have in the creation of peace and in reducing transaction costs. These scholars argue that while regime actors to have self-interests, they are able to see the possibility of creating a global environment were the majority of good can be created for the majority of actors through cooperation. In this approach, no single actor would get the exact regime that it wants, but that through interdependent cooperation it can achieve enough of its aims, while allowing other actors to achieve a sufficient amount of its aims. This approach is designed to create an international regime based on peace and stability.

Those theorists working within the realist/neo-realist and Marxis/neo-Marxist schools tend to focus on the importance of power in the formation and maintenance of international regimes. These global power dynamics can take the form of hegemonic states against weaker ones, or of global power-wielding corporate elite against the unorganized global working class.

Finally, there is a school of regime theory that might be considered post-modernist. Theorists working in this tradition focus on the formation of cognitive frameworks and the ability to set global agendas through the use of media and other tools. These scholars see the regime formation dynamics as based on what forces can influence the acceptable forms of problem definition and solution. These forces form the "epistemic community" for the particular issue are in international affairs, and creates its "accepted" belief system.

While international regime theory provides a very useful theoretical framework to help us understand this period of rapid transformation, there are some problems with its use. In some cases, those that have used regime theory have approached the state as a unitary actor, and ignored domestic contestation to the regime formation processes. Also, in most cases, there is a very heavy focus on state actors, at national, regional and global levels. This focus ignores the increasingly important role played by non-state actors, at each of these levels, particularly by global non-governmental organizations representing the interest of the private sector. Also, there are often insufficient linkages between the processes of global economic restructuring and its influence on domestic actors and political-economic processes. Finally, there is often insufficient attention paid to the factors that affect "state autonomy," or the ability of the state to exercise *de facto* sovereignty.

## Theoretical Framework

In this study, we have primarily adopted the Krasner (1983) approach to international regimes. This causes us to look at the issues of regime transformation, and the emergence of consensus in four critical areas: (1) principles and values; (2) norms; (3) rules; and (4) enforcement mechanisms. However, we also have a focus on the epistemic community, and on the role of global non-state actors in the formation of these regime components.

## Research Questions

<u>Research Question 1</u>: To what degree is the international telecommunications regime eroding, and by what factors is it being eroded?

<u>Research Question 2</u>: What evidence exists that a new regime for information and communications policy is emerging, and what are its principles, values, rules and enforcement mechanisms?

<u>Research Question 3</u>: What are some of the more important regional, sub-regional, and national responses to this transition?

<u>Research Question 4</u>: What are the implications of this regime transformation for developing countries and what can be done to influence the direction of the emerging regime so that it might be more just and equitable for a wider grouping of the world's citizens?

## Methodology and Data

The methodology adopted in this study is qualitative in nature and uses a theory-driven case-study approach. After defining our terms, theoretical framework and research questions, all of which are grounded in the extant literature, multiple qualitative data collection techniques were employed, including: (1) participant-observation; (2) observer-observation; (3) in-depth interviews; and (4) content analysis of primary and secondary sources. Data collection was focused primarily on developing a thick-narrative case study of the impact of regime transformation on an emerging economy. The case selected for analysis was the Republic of South Africa. South Africa was chosen for a number of reasons, including: (1) in 1996 it implemented a fairly wide-ranging restructuring of its telecommunications sector (see, *inter alia*, Cogburn, 1998); (2) it was a founding member of the World Trade Organization; (3) it plays a strong political and economic leadership role in the African region, and within the broader developing country context; (4) it participated in, and made an acceptable offer to the WTO Agreement on Basic Telecommunications; (5) it has developed a merged telecommunications and broadcasting independent regulatory body; (6) it is currently in the process of a Green/White Paper process to develop an electronic commerce policy; and finally (7) there has been significant activity from non-state actors in South Africa, at both the national, regional, and global levels.

## Limitations of the study

Several factors limit the success of this study. The primary factor has been time. The qualitative data collection has yielded a tremendous amount of data and there has been insufficient time for a thorough analysis. Thus, all of the findings and conclusions here should be examined with caution.

A second limiting factor in the study is generalizability. The qualitative paradigm in which this study operates suggests that the primary purpose is not to necessarily generalize the findings from this study to other countries in Africa, or other emerging regions. However, there is a desire to be able to draw some broader conclusions about the emergence of a new GII/GIS regime, and its implications. Unfortunately, the degree to which we will be able to do that from a limited case-study, may be slight.

## Findings and Discussion

### *Technology and Erosion of the Existing Regime*

Are there social, political, economic, cultural, and technological factors that are challenging the International Telecommunications Regime, and if so what are they? The answer to this question is, simply put, yes. Let's look at examples, of each of these factors eroding the International Telecommunications Regime.

Social Factors

Movement towards an information society and many other societal actors getting involved in the process. Emerging out of all of the multiple information society conferences and initiatives discussed above, is a recognition of the many applications that are possible in the emergence of an information and knowledge society (Castells, 1999; Mansell and When, 1999). Table 1.0 below, illustrates the wide range of applications that have been identified for use in a Global Knowledge and Information Society.

**Table 1.0 An Applications Driven Global Information Society**

| General Application | Specific Examples | Epistemic Support |
|---|---|---|
| Education, Research and Training | • Distance-Education<br>• Collaboratories<br>• Asynchronous training | • WTDC<br>• G8 Conference |
| Digital Libraries | • Library of Congress<br>• UMDL<br>• IPL<br>• J-Stor | • WTDC<br>• G8 Conference |
| Electronic Museums and Galleries | • Louvre | • WTDC<br>• G8 Conference |
| Environment Management | • GIS applications | • WTDC<br>• G8 Conference |
| Emergency Management | • EMS | • WTDC<br>• G8 Conference |
| SMMEs, Employment Creation and E-Commerce | • PeopLink<br>• African Crafts Market | • WTDC<br>• G8 Conference |
| Maritime Information | • Early warning systems | • WTDC<br>• G8 Conference |
| Electronic Government Services | • E-passports<br>• Sharing and re-use of records | • WTDC<br>• G8 Conference |
| Debt Management and financial services | • Debt management systems<br>• Electronic bill payment | • WTDC<br>• G8 Conference |
| Tourism | • Hotel and package booking<br>• Promotion and data mining | • WTDC<br>• G8 Conference |
| Health Care | • Tele-medicine<br>• Health Education and Information | • WTDC<br>• G8 Conference |
| Legislation and Legal Services | • Parliament information systems<br>• Legal database access | • WTDC<br>• G8 Conference |
| Transportation of People and Goods | • Transportation system management | • WTDC<br>• G8 Conference |
| Business Development and Trade Efficiency | • Trade promotion<br>• B-2-B e-commerce | • WTDC<br>• G8 Conference |
| Universal Access | • Community Information Centers<br>• Public Internet Terminals | • WTDC<br>• G8 Conference |
| National Systems of Innovation | • Collaboratories<br>• Geographically-distributed research teams | • WTDC<br>• G8 Conference |
| Entertainment and Leisure | • On-line gaming<br>• Adult 0riented material | • WTDC<br>• G8 Conference |

This focus on applications moves us beyond the original "club" of the telecommunications regime, to include teachers, nurses, small businesses, to name but a few new stakeholders. It is this broad stakeholder grouping that makes the social pressures for this new regime so strong.

Political Factors

Further, this move to focus on applications and to include more societal actors, means that more political actors now see a more valuable stake in this emerging regime. While under the international telecommunications regime, there were clear "stovepipe" relationships between the ITU, and national ministries of communications, the emerging regime is going to broaden out the stakeholders. Ministries of health, finance, education, trade & industry, and others, are increasingly clamoring for a increased role in defining their needs for Global Information Infrastructure and the development of a Global Information Society.

As example, take the G8 Information Society and Development (ISAD) conference discussed above. In South Africa, the planning for this important global event started with the Ministry of Foreign Affairs (MOF). Subsequently, as more people became aware of the potential of an Information Society, the Department of Arts, Culture, Science and Technology (DACST) stepped in to play the lead organizing role. However, as the time for the actual conference approached, the Department of Communications (DOC) wrestled control of the activities away from DACST and held center stage during the actual conference (much to the chagrin of DACST, and its political leaders).

Unlike the telecommunications ministries, which are fairly narrowly focused, and were previously conceived to have a narrow group of stakeholders. These ministries are more broadly focused and have a broad group of stakeholders.

In addition to these national political factors, transformation of the international accounting rate mechanism, and the unilateral declaration by the United States that it will pay only a limited amount for telecommunications settlement payments.

Economic Factors

The development of a global information economy fuels the need for corporations to explore the "global option," and engage in geographically disarticulated research, production, distribution and management. Also, the increased focus on the global trade in services as opposed to the global trade in goods – especially for knowledge oriented services and products.

Cultural Factors

Many people are worried about globalization and the information society leading only to the erosion of national cultures, and the "Americanization" (also called the "McDonalidization") of the rest of the world (see Barber, 199x). The emergence of a Global Information Infrastructure and Global Information Society means that there are now tools and mechanisms to promote the preservation of local cultures and to project them globally (e.g., www.si.umich.edu/chico).

Technological Factors

Voice over IP, VSAT, Call-Back systems, etc.

This discussion has helped to highlight the social, economic, political and technological factors eroding the existing International Telecommunications Regime. Now, we will examine the follow-up question, “Is there evidence that a new regime is beginning to emerge in the transformation of the International Telecommunications Regime, and what are its principles, values, and norms?”

*The Emerging GII/GIS Regime and its Principles, Values, and Norms*

Our theoretical framework would suggest that evidence of an emergent regime would come initially from the development of global consensus on the principles, values, and norms around a particular issue area of international affairs. In this case, we are looking for this consensus in the fundamental issues of information infrastructure development and an Information Society.

As discussed above, one area of emerging consensus is that the Information Society should be “applications driven.” Meaning, that as we look at the other areas of more technical issues, they point of putting in hardware, and increasing bandwidth and telecommunications services, is to provide the information infrastructure for applications.

Another area of harmonization is that there should be universal access to high-bandwidth connectivity, services and the applications described above. How to achieve this objective is still the subject of much debate, but that universal access is a target has generated a high-degree of global consensus. One driver of this consensus is the high level of focus on the potential of global electronic commerce (e-commerce) to meet the employment creation demands and economic development objectives of the developing countries (we will explore the implications of this focus in later chapters).

Perhaps the most controversial principle seeking consensus, but seen as most important by many actors, is the liberalization and privatization of telecommunications markets and reciprocal market access for both products and services. In order to promote harmonization around these important principles, a number of high-level meetings have been held (some of these meetings were discussed above).

The Buenos Aires Action Plan (BAAP) that emerged out of the 1st World Telecommunications Development Conference (WTDC) highlighted these points. These points were then taken up at a higher level of consensus in the meeting of Ministers of the Group of Seven Highly-Industrialized Economies (G7).[7]

What emerged from the G7 Ministerial Meeting was the so-called “Brussels Principles” for the Information Society. Table 2.0 below presents these eight high-level principles.

| **Table 2.0 G7 Brussels Principles** |
| --- |
| • Promoting dynamic competition<br>• Encouraging private investment<br>• Defining an adaptable regulatory framework<br>• Providing open access to networks<br><br><u>While</u><br><br>• Ensuring universal provision of an access to services<br>• Promoting equality of opportunity to the citizen<br>• Promoting diversity of content, and<br>• Recognizing the necessity of worldwide co-operation. |

At this meeting of industrialized country leaders, the only major representative of the developing world was Thabo Mbeki, then Deputy President of the Republic of South Africa (now president). In his keynote address, Mbeki argued that there was no way that a “Global” Information Society could be built with the eight countries sitting around the table at the meeting (at the time Russia was a “visiting” member of the G7, now a full member of the G8). He urged the G7 to consider an initiative that would involve a cross-section of the developing world. His challenge was accepted, and after substantial political maneuvering, the G7 countries met, for the first time, in South Africa, along with a cross-section of the developing world (see Cogburn, 1997).

One unstated objective of that meeting was an attempt on the part of the developed countries to further the emerging consensus, embodied in the “Brussels Principles,” within key leadership in the developing world. However, this objective met with substantial resistance, when the developing countries (led by South Africa) refused to simply “endorse” the Brussels Principles as stated. After significant, behind the scenes activity and near crises, a modified statement was accepted, known as the “ISAD Principles.” They included the Brussels Principles, but added several items of particular interest to the developing countries. In some cases these ISAD principles were not new, but simply further clarified (or qualified) aspects of the Brussels Principles to be more “just and equitable.”

[7] The G8 has now taken up the question of a “Digital Divide” as a centerpiece of its strategy by enacting a so-called “DOT Force” – Digital Opportunities Task Force.

| **Table 4.0 ISAD Principles for the Information Society** |
| --- |
| • Universal service<br>• Clear regulatory framework<br>• Employment creation<br>• Global co-operation and competitiveness<br>• Diversity of applications and content<br>• Diversity of language and culture<br>• Co-operation in technology<br>• Private investment and competition<br>• Protection of intellectual property rights<br>• Privacy and data security<br>• Narrowing the infrastructure gap<br>• Co-operation in research and technological development |

Hopefully, this discussion serves to illustrate the point of the multiple processes that have occurred during the inter-regnum between the demise of the International Telecommunications Regime and the emergence of a new Global Information Infrastructure and Global Information Society Regime. There is another interesting story to be told about what happened to the momentum generated by this ISAD conference and its declarations, but that full story is beyond the scope of this study. For the moment, this brief account will have to suffice.

At the close of the ISAD conference, the head of the Egyptian delegation took the floor during the ministerial meeting and urged for there to be a follow-up conference, and pledged Egypt as the host for the conference. This position was completely un-caucused within the developing world delegates. Nonetheless, it was accepted by the ministers, and it was agreed that Egypt would host the follow-up conference the subsequent year. This conference never occurred. Even with substantial urging from the European Commission, the United Nations Economic Commission for Africa (ECA) and other regional and global bodies, no further ISAD activities have been held.

In the vacuum created by the failed ISAD movement, a new initiative led—not by the developing world, as the ISAD processes were—but by the World Bank emerged. This new initiative, called the Global Knowledge for Development conference, was launched in June 1997 in Toronto, Canada. At the inaugural meeting, then South African minister of telecommunications, Jay Naidoo took the microphone in one session and emphatically stated that "this is not ISAD, the ISAD follow-up meeting is still to come." But alas, it was not.

To summarize, table 2.0 below presents a comparison of the "old" International Telecommunications Regime and the "emerging" GII/GIS Regime. Table 3.0 that follows presents a brief chronology of these major regime transformation events.

| Table 2.0 Comparison Between ITR and Emerging GII/GIS Regime | |
|---|---|
| **International Telecommunications Regime** | **Emerging GII/GIS Regime** |
| Limited Competition: Natural Monopoly for Telecommunications | High Competition: Liberalization and privatization for telecommunications |
| Single Issue: Telecommunications | Multiple Issue: Telecoms, broadcasting, health, education, SMMEs, debt magangment, etc. |
| Single Ministry: Telecoms, PTT | Multiple Ministries: Broadcasting, Education, Health, Trade & Industry, Finance, etc. |
| Single Industry: Telecoms and equipment suppliers | Multiple industries: Content providers, ASPs, ISPs, e-commerce, etc. |
| Limited Stakeholders: Telecoms employees and experts | Multiple Stakeholders: Nurses, educators, small-business owners, etc. |
| Epistemic Community: Narrow | Epistemic Community: Wide |

| 3.0 Brief Chronology of Regime Transformation | |
|---|---|
| • WTO Born (January 1995)<br>• G7 Ministerial Meeting<br>• GII Principles (February 1995)<br>• GIIC Formed (February 1995)nWTO Financial Services Agreement (July 1995)<br>• WTO Agreement on Movement of Natural Persons (July 1995)<br>• ITU Telecom (October 1995)<br>• ISAD (May 1996)<br>• AISI Adopted (May 1996)<br>• WTO, IMF World Bank MOU (November 1996)<br>• WTO Agreement on Basic Telecoms (February 1997) | • WTO Agreement on IT Products (March 1997)<br>• Global Knowledge for Development (June 1997)<br>• US Issues Unilateral Challenge to the International System of Accounting Rates at ITU (1997)<br>• WTO E-Commerce Work Programme (September 1998)<br>• OECD E-Commerce Ministerial Conference (October 1998)<br>• WTO Seattle Ministerial Meeting (1999)<br>• Washington, D.C. Meeting of the IMF and World Bank (1999)<br>• World Economic Forum meeting in Davos (2000) |

So we can clearly see the contours of principles, values and norms emerging for a new Information Infrastructure Regime. However, this transition to a new regime will not be completely smooth. Already, there has been significant contestation to the emergent regime. There has been a contest between the United States and the European Commission over privacy concerns, especially related to Global E-Commerce. Here, we see a fundamental clash of principles and values. The European perspective is based on a cherished belief that individuals have the right to own information about themselves, and should give authorization for any information to be collected, stored, or disseminated.  In contrast, the dominant U.S. perspective is that companies have the right to collect information on their citizens, in order to be able to provide them with better and more targeted advertising and services.

Significant opposition is emerging to globalization, and global process of developing regime principles, norms and rules by elite bodies, behind closed doors, regardless of whether or not those doors are in Seattle, Washington, D.C. or Davos.   There was significant hesitation by developing countries to make offers to the WTO Agreement on Basic Telecommunications (for example, in the end, only five African countries – including South Africa – actually made accepted offers, thus joining the agreement). Further, even South Africa has refused to join the WTO Information Technology Agreement, which will lower tariffs on imports of mostly all information technology products to near zero.

*Rules and Enforcement of a New GII/GIS Regime*

If we now agree that the old regime is being transformed, and a new regime is emerging, complete with its own set of principles, values, and norms.  What are the rules of this newly emerging regime and what international body will enforce these rules?

To answer the second question first, there is little doubt that the centerpiece organization will be the World Trade Organization.  However, unlike the International Telecommunications Regime that was based primarily on a single intergovernmental organization, the ITU, the emerging regime will rely on a host of governmental and non-governmental organizations to enforce its rules.

Thus, in addition to the WTO, the ten most important organizations for the "governance" of this emerging regime will be the following: (1) World Intellectual Property Organization; (2) Organization for Economic Cooperation and Development; (3) Internet Corporation for Assigned Names and Numbers; (4) Global Information Infrastructure Commission/Global Business Dialogue; (5) Group of 8 Industrialized Countries; (6) World Economic Forum; (7) World Bank Group; (8) European Commission; (9) International Telecommunication Union; and (10) Bi-lateral aid agencies.  Table 5.0 illustrates these organizations, their organizational type, and the primary regime function.

| Table 5.0 GII/GIS Regime Enforcement Organizations | | |
|---|---|---|
| **Organization** | **Organization Type** | **Regime Component(s)** |
| WTO | Inter-Governmental (Global) | Principles, Values, Norms, Rules, **Enforcement** |
| WIPO | Inter-Governmental (Global) | Principles, Values, Norms, Rules, **Enforcement** |
| OECD | Inter-Governmental (Regional) | Principles, Values, Norms |
| ICANN | Global Non-Governmental | Principles, Values, Norms, Rules, **Enforcement** |
| GIIC/GBD | Global Non-Governmental | Principles, Values, Norms |
| G8 | Inter-Governmental | Principles, Values, Norms |
| WEF | Global Non-Governmental | Principles, Values, Norms |
| World Bank Group | Inter-Governmental (Global) | Principles, Values, Norms, Rules, **Enforcement** |
| European Commission | Inter-Governmental (Regional) | Principles, Values, Norms |
| ITU | Inter-Governmental (Global) | Principles, Values, Norms |
| Bi-Lateral Aid Agencies | Governmental | Principles, Values, Norms, Rules, **Enforcement** |

To illustrate this point, take the example of South Africa, as it attempts to develop its new e-commerce policy. What are the on-going processes of global regime formation within the international system that have an impact on South Africa's ability to harness the digital economy? What has been South Africa's recent involvement (1995-2000) in these on-going processes of international regime formation?

International Regime Theory and the Digital Economy

The transition to a global information-oriented economy has created new challenges for global governance and regulation of these processes. The borderless nature of this global economy raises fundamental questions about how we might be able to achieve higher levels of harmony and reduced transaction costs as electronic commerce continues to develop. These fundamentally global processes are illustrative of a quandary facing international relations theorists around the world: When dealing with issues that are transnational in scope, how do we achieve significant levels of global governance in the absence of a global government? This dilemma, often referred to by scholars as the "anarchy problematique," has confronted numerous issue areas from diamonds to telecommunications to the high-seas. One solution to this dilemma comes in the form of international regime theory, where we try to look at the emergence of norms, principles and values around a particular issue area, and mechanisms to enforce them.[8]

[8] For theoretical background on international regime theory see, *inter alia*, *International Organization*. For empirical application of regime theory, see, *inter alia*, Peter Cowhey, "Roots of High-Technology Regimes: The International Telecommunications Regime."

Globally, there are a number of primary norms, principles and values that are emerging around electronic commerce and the information economy.[9] Some of these include: (1) telecommunications and information infrastructure – the importance of liberalization, privatization and a pro-competitive environment; (2) customs/taxation – that the Internet and e-commerce should continue to be a "tax free" zone; (3) electronic payments – that multiple options should continue to emerge (both inside and outside money) that is interoperable and allowing for both anonymous, pseudonymous, and traceable methods; (4) commercial code – that a common global commercial code should emerge to provide for the global rule of law and protection for contracts and private property; (5) intellectual property protection – that IPR regulation needs to be revised to reflect the realities of the digital economy, while still providing an incentive for the production of information goods; (6) that domain names – are an important and contested commercial asset, and famous marks should be protected while not allowing them to abuse smaller enterprises, and that ICANN is the legitimate body charged with the responsibility to deal with domain name issues; (7) personal data – should be protected, while at the same time allowing for legitimate corporate uses of data profiling and targeted advertising; (8) security and encryption – is an important national and personal security concern that has to be balanced with personal privacy concerns; (9) awareness/trust – is a limiting factor for the growth of e-commerce; (10) trust – might be enhanced with the widespread use of authentication and digital signatures; (11) technical standards – should be technology neutral and industry driven to the fullest extent possible; (12) local content – should be promoted and protected, if e-commerce is going to reach its full potential; (13) labor and society – will be affected by the move towards a digital economy and we should work to minimize the negative impact, while harnessing the potential; (14) universal service/access – or lack thereof, as characterized by the "digital divide" is one of the most potentially limiting factors for global e-commerce, and finally (15) human resources and capacity – require immediate global attention.

One of the most successful international regimes in history was the International Telecommunications Regime.[10] This regime, was based primarily on a specialized agency of the United Nations, the International Telecommunication Union (ITU). However, due to various social, political, economic, and technological factors, this regime is being eroded and new regimes are emerging, the broadest of which might be called the Global Information Infrastructure/Global Information Soceity (GII/GIS) Regime an critically important sub-set of which is the Global Electronic Commerce Regime.[11]

Unlike the highly successful international telecommunications regime, which was based primarily on one intergovernmental organization, the norms, principles and values of the emergent Information Regime are being promoted, debated, contested, and will ultimate be enforced by a range of global organizations, both intergovernmental and non-governmental. At the center of this regime's enforcement structure is the World Trade Organization (WTO), supported by the ITU and various specialized international and regional organizations, such as the World Intellectual Property Organization (WIPO), the Organization for Economic Cooperation and Development (OECD), International Labor Organization (ILO), International Organization for Standards (ISO), the Internet Corporation for Assigned Names and Numbers (ICANN), United Nations Education, Scientific and Cultural Organization (UNESCO), United Nations Conference on Trade and Development (UNCTAD) and the World Bank. In addition, several national governments and regional governmental groupings have taken important steps towards influencing the emerging regime, such as the Group of Eight Industrialized Nations (G8), European Commission (EC), Government of the United States of America (USG), and the Asia Pacific Economic Cooperation (APEC). Each of these organizations will play a specific and important role in providing governance to the international regime for global electronic commerce and the Information Society.

---

[9] Derrick L. Cogburn, "Transformation and Emergence of a Global Information Regime," *The Emergent Information* Regime, forthcoming, 2001.

[10] Peter Cowhey, "The International Telecommunications Regime," *International Organization.*

[11] Derrick L.Cogburn, "Global Governance in the Information Age: Regime Erosion, Transformation and Emergence," *The Emergent Information Regime*, forthcoming.

The Importance of Policy Issues

While technical issues are critical to the development of electronic commerce and the digital economy, they are far from decisive. A number of other legal and regulatory hurdles that have to be addressed. Jonathan Coppel of the OECD argues that

Despite the phenomenal growth in the Internet for commercial purposes there are a number of legal and technical obstacles which could hinder the full potential of e-commerce from being reaped. For example, the virtual environment of electronic markets makes it more difficult to determine who the contracting parties are, where an electronic commerce operator is established and whether that operator is complying with all relevant legal obligations and regulatory regimes .... And the absence of commercial codes and legal recognition covering areas such as the acceptance of electronic signatures and documents, contrat enforcement and greater certainty vis-à-vis liability for damages that may arise as a result of electronic transactions, will limit the take-up of e-commerce, particularly in the B2B sphere. These concerns are magnified when trading across borders. (OECD 2000).

> "..the virtual environment of electronic markets makes it more difficult to determine who the contracting parties are, where an electronic commerce operator is established and whether that operator is complying with all relevant legal obligations and regulatory regimes".

Further concerns are raised about electronic transactions across border, for example, the difficulty of verifying electronic signatures and documents.

Progress towards and e-commerce regime in South Africa

While global e-commerce is being driven in many ways by the leadership of the private sector, there are very important information policy issues that will facilitate its optimal growth, both within South Africa and around the world. One challenging paradox of e-commerce is that while its scope is clearly global, national regulation continues to provide the legal and regulatory basis for its operation.

The South African government has taken these responsibilities very seriously, and the Department of Communications has launched an important national Green/White paper process on electronic commerce that will lead to specific national legislation by the 3rd or 4th quarter of 2001.[12] The process of developing and conducting this process has been consultative and tried to include the voices of as many relevant stakeholders as possible.[13] This section will briefly examine the policy perspectives that are emerging in South Africa's movement towards and e-commerce regime. Our primary data source for this section is the national Green Paper on Electronic Commerce, the background papers commissioned by the Department of Communications, the papers of the working groups, other published government documents, academic literature, and news accounts.

Of significant interest for our analysis, is the fact that the Green Paper makes constant reference to the need to harmonize its emerging national e-commerce regime with the growing global consensus and in line with its extant commitments to the World Trade Organization. "In embarking on a national policy development initiative on e-commerce it is imperative that SA take cognizance of its WTO commitments, firstly, to ensure that such policy is compatible with the relevant WTO rules and regulations, and secondly, to determine the impact of e-commerce on those commitments."[14] The WTO has worked to review the impact of e-commerce on its structure and planning. At its last Ministerial Meeting, held in Seattle, Washington, the US and other developed countries, wanted to explore the possibilities of a more comprehensive involvement for the WTO in e-commerce issues. "In the Seattle Ministerial Conference, South Africa, together with the Southern African Development Community (SADC), supported the extension of the moratorium until the next Ministerial Conference when it would be reviewed."[15] The current policy perspective recognizes that "any regulatory regime that South Africa adopts must be consistent and compatible with international frameworks."[16]

General Principles on Electronic Commerce – Consensus on general principles around issues of international import are a key indicator of the emergence of a new regime. In terms of the e-commerce policy formulation process, South Africa's approach is based on eight key principles, which are: (1) quality of life; (2) international benchmarking; (3) consultative process; (4) flexibility; (5) technology neutrality; (6) supporting private-sector led and technology-based solutions and initiatives; (7) establishing and supporting public-private partnerships, and supporting small, medium and miro-sized enterprises (SMMEs).[17]

---

[12] http://www.polity.org.za/govdocs/green_papers/greenpaper/index.html
[13] http://www.isoc.org/inet99/proceedings/1g/1g_4.htm
[14] http://www.polity.org.za/govdocs/green_papers/greenpaper/index.html, p. 48.
[15] http://www.polity.org.za/govdocs/green_papers/greenpaper/index.html, p. 49.
[16] http://www.polity.org.za/govdocs/green_papers/greenpaper/index.html, p. 18.
[17] http://www.polity.org.za/govdocs/green_papers/greenpaper/index.html, p. 18.

In terms of the substantive principles, South Africa believes the following: (1) the recognition that there is a need for legislation to support the national implementation of e-commerce transactions, within a framework of international standards; (2) that commercial transactions should be able to be effected through both paper and electronic means, without creating uncertainty about the latter; (3) promoting a framework that increases the efficiency of South African commercial transactions, without being overly cumbersome; (4) the framework should be technology neutral; (5) to develop a uniform commercial framework that conforms to international standards; (6) that South Africa should build on the work of others and not reinvent the wheel; and that (7) South Africa should strive to maintain its sovereignty and independence, and meet its strategic national socioeconomic development objectives.[18]

Telecommunications and Information Infrastructure – Without increased access to information and communications infrastructure, e-commerce will not be able to meet its full potential.[19] Since the restructuring of the telecommunications sector in South Africa in 1996, there have been a number of information infrastructure initiatives in the country (Cogburn 1998). The Department of Communications (DoC) has been at the forefront of this effort, particularly with its *Info.Com 2025* Strategy, Public Information Terminals, Public Key Infrastructure Pilot, and numerous other e-commerce and e-government initiatives. As these infrastructure initiatives unfold, the strategy should be to develop and infrastructure that is capable of handling a wide variety of applications and services. From the South African policy perspective, "the challenge confronting South Africa is to create an ideal market structure for e-commerce that will stimulate and modernise network development and infrastructure; accelerate universal access; support affordable access; encourage investment and innovation."[20] There is a realization in the Green Paper that the infrastructure for e-commerce will consist of a range of networks, including "backbone networks, end-user equipment and access services."

> The success of e-commerce will depend on the available of speedy access infrastructure; high quality of service within the backbone networks; and affordable prices. Access will not only be through fixed networks (terrestrial, wireline and cable TV) but also through wireless networks (cellular, satellite, and digital broadcast spectrum).[21]

Perhaps one of the most important emerging regime principles is the importance of liberalization, privatization and a pro-competitive environment for telecommunications and information infrastructure. South Africa is proudly a founding member of the World Trade Organization (WTO), and has been working actively to promote the multilateral trading system.[22]

---

[18] http://www.polity.org.za/govdocs/green_papers/greenpaper/index.html, pp. 25-28.
[19] http://www.polity.org.za/govdocs/green_papers/greenpaper/index.html, pg. 85.
[20] http://www.polity.org.za/govdocs/green_papers/greenpaper/index.html, pg. 82.
[21] http://www.polity.org.za/govdocs/green_papers/greenpaper/index.html, pg. 83.
[22] Address by Minister of Finance, Trevor Manuel, as chairman of the Board of Governors International Monetary Fund.

At the moment, Telkom, the commercialized Public Telecommunications Operator (PTO) has a monopoly on the provision of basic fixed telephony services. While the government chose to adopt this strategic equity partnership (SBC and Telkom Malaysia) for Telkom, the Green Paper recognizes that "Telkom's efforts alone are not sufficient to achieve all of the infrastructure needs for e-commerce [in South Africa]."[23] As such, South Africa submitted an accepted offer in the WTO's Agreement on Basic Telecommunications, and is now bound by the terms of that agreement to liberalize and privatize its telecommunications sector by 2002.[24] However, at present, South Africa has not yet signed the WTO Information Technology Agreement (ITA), which would bring tariffs on a wide range of information and communications technologies down to zero by 2001.

Universal service/access – As stated above, there is a significant recognition that all of the potential benefits of global electronic commerce for South Africa will not be realized without sufficient attention to increased access to information and communications technologies for a wider portion of South African society. Often characterized as the "digital divide," this disparity of access both within countries and between them is one of the most potentially limiting factors for global e-commerce.

In order to combat the digital divide and try to meet its universal service goals, the Department of Communications has promoted a number of public access initiatives such as the development of Multi-Purpose Community Information Centers (MPCICs), the Universal Service Agency (USA), and Public Information Terminals (PITs) to help South Africa to reach provide access for larger numbers of its citizens to the benefits and opportunities of global electronic commerce.

Customs/Taxation – South Africa recognizes that the transition to a digital economy engenders new ways of doing business, and new products and services. Many of these products and services are presenting tremendous challenges to the taxation regimes of governments around the world. "There is a legitimate concern by certain governments that the development of the Internet may have the effect of shrinking the tax base and hence reducing fiscal revenue."[25] I n addition, South Africa recognizes that there are significant difficulties in defining jurisdiction in electronic commerce, and to administer and enforce any kind of taxation scheme.

The South African Revenue Service (SARS) believes that the global consensus that is emerging around taxation principles, being led by the Organization for Economic Cooperation and Development (OECD), does not conflict with its views. The important basic principles of this emerging regime are: (1) neutrality; (2) efficiency; (3) certainty and simplicity; and (4) flexibility. Of particular interest, there is apparently no opposition in the South African approach to the idea of "no need for a special new tax such as a "flat rate" or a "bit" tax, and that the Internet and e-commerce should continue to be a "tax free" zone.[26]

---

[23] http://www.polity.org.za/govdocs/green_papers/greenpaper/index.html, pg. 85.
[24] WTO Agreement on Basic Telecommunications.
[25] http://www.polity.org.za/govdocs/green_papers/greenpaper/index.html, p. 36.
[26] http://www.polity.org.za/govdocs/green_papers/greenpaper/index.html, p. 37.

However, South Africa wants to promote the idea of "indirect taxes", being at the place of consumption. "Indirect taxes should apply where consumption taxes place, and an international consensus should be sought on the identification of the place of consumption. Consensus is important to avoid double taxation or unintentional non-taxation." [27]

There is concern in South Africa that the development of electronic money that is "unaccounted", and "network" or "outside" money, will lead to additional challenges in terms of tax monitoring, collection and enforcement. However, it believes that there is significant cultural conservatism that will limit the impact of these new forms of money. In order to promote compliance, South Africa believes that it should require that certain information should be a part of South African e-commerce.

> The following information should be furnished on any commercial website owned by a South African resident, company, close corporation or trust: trading name of the business; the physical as well as the postal address for the business; and e-mail address; telephone or other contact information and statutory registration number in respect of companies; close corporations and trusts.[28]

The emerging tax perspective recognizes that there are additional complications that reduce storage and transmission costs, and that storing information overseas is becoming easier and cheaper. As a result, South Africa believes that there is the need for a "greater degree of international co-operation in revenue collection than currently exists."[29] It appears that South Africa supports the role of the OECD, as a leader for this aspect of the regime, especially with its Model Tax conventions.
Electronic payments – The emerging policy perspective in South Africa is that multiple options should continue to emerge (both inside and outside money) that are interoperable and could allow for both anonymous, pseudonymous, and traceable methods. There is particular concern about the "threat of cybercash" and the impact of unaccounted money on the South African economy (both in the form of network-based money and stored value cards). Both of these methods have the potential to exchange value without identifying the user and without linking to specific bank accounts.[30] South Africa sees this as a "make-or-break" issue for electronic commerce in South Africa.

Another major challenge for South Africa, given its history of racial oppression and segregation, is the ability for the "unbanked" have access to electronic payment systems.

South Africa has a well-developed financial system, and the South African Reserve Bank (SARB) has taken the lead on these e-payment issues. In 1998, it developed the South African Multiple Option Settlement (SAMOS) system that allows real-time settlement between banks. The SARB has also published a position paper on e-money in April 1999. The Reserve Bank is pushing hard for the principle that "only banks would be allowed to issue electronic money," although there is the recognition that "the issuance of electronic money may fall outside the definition of [the] 'business of a bank', as defined in the Banks Act 94 of 1990.[31] The goal is to protect users, who the Reserve Bank feels may find themselves "unprotected," in the event that the issuers of electronic money remain unregulated. The Reserve Bank feels strongly that "primary and intermediary issuers of electronic value will therefore be subject to regulation and supervision by the South African Reserve Bank."[32]

[27] http://www.polity.org.za/govdocs/green_papers/greenpaper/index.html, p. 40.
[28] http://www.polity.org.za/govdocs/green_papers/greenpaper/index.html, p. 44.
[29] http://www.polity.org.za/govdocs/green_papers/greenpaper/index.html, p. 45.
[30] http://www.polity.org.za/govdocs/green_papers/greenpaper/index.html, p. 99.

<u>Global Commercial Code</u> – South Africa recognizes that global electronic commerce is posing a challenge to its national legal systems that support commercial transactions. The current legal framework in South Africa, like in most countries, was developed for an era of paper-based commerce, and thus contains words such as: "document", "writing," " signature," "original," "copy," "stamp," "seal," "register," "file," "deliver," etc.[33] The South African Law Commission found that the Computer Evidence Act 57 of 1983 was insufficient to address the admissibility of "computer evidence" in civil proceedings, and this will have to be addressed in a emerging e-commerce regime.

Also important is the ability to determine the attribution of electronic documents. Given the existing law in South Africa, this issue has to be addressed.

> However, in terms of the doctrine of 'estoppel' in South African law, a purported originator who never sent nor authorized a communication to be sent, may nevertheless be held bound in law if his negligent conduct, whether by action or commission, induced a reasonable belief of authenticity in the mind of the addressee, which caused the latter to act thereon to his/her peril.[34]

Additionally, it is important to ascertain the time and place of an e-commerce contract, in order to determine whether or not South African courts have "jurisdiction to adjudicate a dispute involving both local and foreign nationals and, if so, which country's laws our courts would apply."[35] How to effect a signature in cyberspace is another important issue for the South African policy environment. A framework for understanding electronic signatures (and the more specific subset "digital signatures") must be put into place, and a common global commercial code should emerge to provide for the global rule of law and protection for contracts and private property.

As the leading regime component in this area, South Africa strongly supports the United Nations Conference on International Trade Law (UNCITRAL) and its Model Law on Electronic Commermce.

---

[31] http://www.polity.org.za/govdocs/green_papers/greenpaper/index.html, pg. 102.
[32] http://www.polity.org.za/govdocs/green_papers/greenpaper/index.html, pg. 102.
[33] http://www.polity.org.za/govdocs/green_papers/greenpaper/index.html, p. 28.
[34] http://www.polity.org.za/govdocs/green_papers/greenpaper/index.html, p. 32.
[35] http://www.polity.org.za/govdocs/green_papers/greenpaper/index.html, p. 32.

Intellectual Property Protection – The South African policy approach recognizes that the transition to a digital economy presents new challenges for intellectual property protection. Digital goods can be copied and distributed around the world with relative ease, putting additional pressure on the system of intellectual property protection in South Africa, and countries around the world. Intellectual property regulation needs to be revised to reflect the realities of the digital economy, while still providing an incentive for the production of information goods, and thus balancing the needs of the individual with the needs of society.

"South African intellectual property law is not fully equipped to deal with the implications of the Internet, convergence, multimedia, digital technology and hence ecommerce. The advent of the Internet has changed the underlying assumptions of the original copyright laws entailed in the Copyrights Act 98 of 1978."[36] South Africa has already made an attempt to comply with the WTO's Agreement on the Trade-Related Aspects of Intellectual Property (TRIPS) by amending its Intellectual Property Laws Amendment Act (Act 38 of 1997).

In order to try to help move forward the development of a global e-commerce regime, the Word Intellectual Property Organization (WIPO) has developed its "digital agenda" to guide its work in this area over the course of the next two years. The South African Department of Trade and Industry (DTI) convened a consultative meeting in South Africa to discuss South Africa's accession to these WIPO treaties and processes. "The majority of stakeholders cautioned that before acceding to them, South Africa should analyse the benefits which accrue to small and medium enterprises."[37]

Domain Names – Currently, there are no direct linkages between domain names and trademark holders. This area, perhaps better than any other, highlights the significant contradictions that are at play in the development of global electronic commerce, en an environment of national-based legislation. As the South African *Green Paper* argues that: "Trademarks are territorial in nature, i.e. their registration applies to a particular country or jurisdiction. There is a general discrepancy between the national scope of trademark and the international nature of electronic commerce, particularly since e-commerce is borderless and instantaneous in nature."[38]

South Africa recognizes that domain names are an important and contested commercial asset, and famous marks should be protected while not allowing them to abuse smaller enterprises. There is some concern that ICANN has not yet achieve complete legitimacy as the body charged with the responsibility to deal with domain name issues. South Africa is questioning whether or not it should support these structures, as well as structures such as AfriNIC, which has been formed to try to better represent the interests of Africa within ICANN.[39]

---

[36] http://www.polity.org.za/govdocs/green_papers/greenpaper/index.html, p. 57.
[37] http://www.polity.org.za/govdocs/green_papers/greenpaper/index.html, p. 60.
[38] http://www.polity.org.za/govdocs/green_papers/greenpaper/index.html, p. 63.
[39] http://www.polity.org.za/govdocs/green_papers/greenpaper/index.html, p. 97.

South Africa does, however, support the role of WIPO in its dispute resolution activities. It also supports the idea that in an information economy, the so-called Country Code Top Level Domains (ccTLDs) should be managed by national governments as a national asset.[40] The South African Department of Communications has proposed the creation of an Independent Domain Name Authority (DNA) to represent all relevant stakeholders (private sector, public sector, and civil society) and to manage the domain name issues for South Africa.

Personal Data and Consumer Protection – In order to enhance trust in the digital economy, South Africa recognizes that personal date should be protected. The challenge is to what degree the South African policy perspective will allow for legitimate corporate uses of data mining and profiling, targeted advertising, and the use of other Customer Relationship Management (CRM) tools. As fundamental principles, South Africa believes that consumers should be protected against the following dangers:

- Unsolicited goods and communication;
- Illegal or harmful goods, services and content (e.g. pornographic material)
- Dangers resulting from the ease and convenience of buying on-line;
- Insufficient information about goods or about their supplier; since, the buyer is not in a position to physically examine the goods offered;
- The abundantly accessible nature of a website;
- The dangers of invasion of privacy;
- The risk of being deprived of protectin through the unfamiliar, inadwauate or conflicting law of a foreign country being applicable to the contract, and finally
- Cyber fraud.[41]

South Africa also recognizes that when moving into electronic commerce, suppliers also face new dangers, especially in exposing themselves to new liabilities. The South African policy process would like to ensure that South African digital enterprises are an attractive competitor in the cyber world. The Department of Communications sees this as "an opportunity [for South African businesses] to establish a reputation for sound e-commercial practices, not only locally or within the SADC but also worldwide."[42]

Of particular importance to South Africa is the impact that its privacy and consumer protection policies may have on its relationships with its trading partners, especially the European Union which has a very stringent privacy policy and consumer protection perspectives. There is a recommendation in the *Green Paper* that "a combined government and industry database be set up to enable South African businesses to establish practices in any EU member country from which they may acquire personal data, for example, to establish profiles of their customers in that country."[43]

---

[40] http://www.polity.org.za/govdocs/green_papers/greenpaper/index.html, p. 94.
[41] http://www.polity.org.za/govdocs/green_papers/greenpaper/index.html, p. 75.
[42] http://www.polity.org.za/govdocs/green_papers/greenpaper/index.html, p. 78.
[43] http://www.polity.org.za/govdocs/green_papers/greenpaper/index.html, p. 80.

Security, Encryption and Trust – South African believes that "security measures used in conventional commerce may not be adequate to provide trust in the electronic economy."[44] At the same time, it is important that national and personal security concerns are balanced with personal privacy concerns. Four key elements are seen as crucial to ensuring that transactions in the digital economy can take place securely. These elements are: (1) authentication; (2) confidentiality; (3) integrity; and (4) non-repudiation. From South Africa's perspective, achieving this level of security for the digital economy "requires active partnership between government and the private sector."[45]

These technologies are seen as critical to promoting trust in the digital economy, amongst both consumers and producers. It appears that South Africa is comfortable with the leading role being played by the Organization for Economic Cooperation and Development (OECD) in promoting a regime consensus in this area.

Awareness – In South Africa, as in many other parts of the world, low levels of awareness about the potential benefits and opportunities in electronic commerce, is a limiting factor for its growth. South Africa is developing a strategy to promote these opportunities, both to consumers and amongst the SMME sector.

> Central to this issue is educating the wider population about both the opportunities and potential, threats of e-commerce. Coupled with that is the need to popularize or publicise and e-commerce policy process so as to invite participation. The creation of awareness and other related initiatives by government and its partners from the academic and business sectors to promote technological development should be done on an integrated approach. We need to build a new e-community that can take effective advantage of the e-commerce opportunities.[46]

Within the South African public, private and civil society sectors, there are many bodies working to promote this level of awareness. Within the government, the Department of Communications is playing a leading role. Numerous private sector enterprises and bodies such as the Electronic Commerce Association of South Africa (ECASA) and the African Connection are also contributing in this area. In the civil society, the University of the Watersrand's Learning, Information, Networks, and Knowledge (LINK) Center is engaged in promoting an enhanced intellectual understanding of these issues, and the Internet Society of South Africa is building technical and user awareness.

Technical standards – The emerging South African perspective on technical standards is that they of critical importance to the development and proper functioning of the Internet and global electronic commerce. "Standards are rules, and serve as a basis for comparison and a form of order. The major objective for standardization is to achieve interoperability between networks and services and ensure compatibility."[47]

---

[44] http://www.polity.org.za/govdocs/green_papers/greenpaper/index.html, p. 66.
[45] http://www.polity.org.za/govdocs/green_papers/greenpaper/index.html, p. 66.
[46] http://www.polity.org.za/govdocs/green_papers/greenpaper/index.html, p. 112.
[47] http://www.polity.org.za/govdocs/green_papers/greenpaper/index.html, p. 91.

"Standards are needed for long-term commercial success of the Internet since they can allow products, services and applications from different firms to work hand in hand. Standards encourage competition and reduce stress or uncertainty in the market place."[48] However, there is also a recognition that "Standards can also be employed as de-facto non-tariff trade barriers to "lockout' non-indigenous business from a particular national market."[49]

Further, there are also the tremendous challenges of developing standards in "an environment in which technology is developing rapidly may be counter productive at this stage of e-commerce."[50] There is the recognition that these standards should be technology neutral and industry driven to the fullest extent possible.
South Africa supports the international organizations playing the leading role in developing this component of the global e-commerce regime, especially the role of the International Standards Organization (ISO) and the International Telecommunication Union (ITU).[51]

Local Content – There are numerous possibilities for promoting local content in the digital economy. In South Africa, there is a growing recognition that perhaps the primary source of this local content will be the growth and development of the SMMEs sector. Small, medium and micro-sized enterprises will be looked to increasingly to create employment opportunities for South Africa.

Several international organizations both governmental and non-governmental, including the United Nations Conference on Trade and Development (UNCTAD), the World Intellectual Property Organization (WIPO), the International Chambers of Commerce (ICC) and others are working to promote the impact of both developing countries and SMMEs on the digital economy.

Labor and Society – As South Africa moves towards a digital economy, it is important to work to minimize the negative impact of e-commerce, while harnessing its potential. It is clear that both of these aspects are real possibilities in South Africa. On the one hand, new growth and new types of employment are indeed possible, while on the other hand, "many workers could be come displaced, temporarily or permanently as a result of this transformation."[52] "Clearly there is need for research in this area to evaluate the nature and number of jobs that could be created by e-commerce and lost or displaced due to efficiencies brought about by new ways of doing business and consumers, a new breed of e-commerce firm "the infomediary" is being created to exploit the Internet."[53]

---

[48] http://www.polity.org.za/govdocs/green_papers/greenpaper/index.html, p. 91.
[49] http://www.polity.org.za/govdocs/green_papers/greenpaper/index.html, p. 92.
[50] http://www.polity.org.za/govdocs/green_papers/greenpaper/index.html, p. 92.
[51] http://www.polity.org.za/govdocs/green_papers/greenpaper/index.html, p. 92.
[52] http://www.polity.org.za/govdocs/green_papers/greenpaper/index.html, p. 112.
[53] http://www.polity.org.za/govdocs/green_papers/greenpaper/index.html, p. 112.

Globally, many of the high-technology workers that have sought fame and fortune in the digital economy, are now becoming highly disillusioned.[54] Recently, high technology workers at one of the most widely know e-commerce companies, Amazon.com, have attempted to unionize in the Washington Area Technology Workers (WashTech), a union structure within the Communications Workers of America (CWA).

Currently, the International Labor Organization (ILO) is reasserting itself as an important player in the international regime formation process for e-commerce, with a focus on understanding the impact on labor issues.

Human resources and capacity – While the shortage of human resources with the requisite skills in information and communications technologies requires immediate global attention, this situation is particularly problematic in South Africa. In South Africa, the Human Science Research Council (HSRC) states that, "there is a chronic shortage of highly skilled human resources in various segments of the market. The scarcity of technical expertise and skills, in the country is further exacerbated by the "brain drain".[55]

South Africa further recognizes that human development must occur on at least five different levels: (1) skills and human resources; (2) digital literacy; (3) digital skills for all South Africans; (4) skills for business; and (3) skills for the future.[56] Distance education and virtual campuses are seen as important elements of this strategy, and should be supported and developed in South Africa.

*Contesting and Shaping the Emerging Regime*

As we can see from this case study of South Africa, this newly emerging regime will be wide ranging, and have a tremendous impact on nearly every area of how we "live, work, and play," as the evolving mantra goes. If this is so, it means that this regime involves perhaps the most important set of principles, norms, and values that we have seen in an international regime. What can be done to influence the direction of the emerging regime so that it might be more just and equitable for a wider grouping of the world's citizens?

## Conclusions and Future Research

This has been a wide ranging study of a very important topic. For me, it is of more than academic importance that we understand this historical period of globalization and the Information Society, and even more so that we understand the erosion of the old International Telecommunications Regime and the emergence of a new form of global governance. The importance is increased, as we grasp how wide-ranging the impact of this regime will be.

---

[54] *Net Slaves...*
[55] http://www.polity.org.za/govdocs/green_papers/greenpaper/index.html, p. 111.
[56] http://www.polity.org.za/govdocs/green_papers/greenpaper/index.html, p. 111.